\documentclass[preprint,pre,aps,amsfonts,amsmath,superscriptaddress,nofootinbib,showkeys]{revtex4}
\usepackage{amsfonts,amsmath,amsthm,amssymb,amscd}
\usepackage{mathrsfs}
\usepackage{empheq}
\usepackage{graphicx}
\usepackage{bm,bbold}
\usepackage{color}
\usepackage{calligra}
\usepackage[T1]{fontenc}
\newtheorem{condition}{Condition}

\begin{document}
%
%%%%%%%%%%%%%%%%%%%%
%
\title{Diffusion Effects on a Stochastic Amplifier Driven by the Square of a Gaussian Field: Discontinuous Transition to Intermittency}
\author{Philippe Mounaix}
\email{philippe.mounaix@polytechnique.edu}
\affiliation{CPHT, CNRS, \'Ecole
polytechnique, Institut Polytechnique de Paris, 91120 Palaiseau, France.}
\date{\today}
\begin{abstract}
We study the effects of diffusion on the transition to intermittency in stochastic amplification driven by the square of a Gaussian field. The transition is characterized by an order parameter quantifying the spatial uniformity of the amplification. We consider a model in which the diffusion-amplification process is embedded in a random lattice of point absorbers in such a way that the Feynman-Kac formula is restricted to Brownian paths with a range not exceeding a finite cut-off. In this setting, with the Gaussian driving field in a wide class of physical interest, we show that diffusion-induced damping of the amplification is sufficiently strong to change the nature of the transition: from continuous in the absence of diffusion, to discontinuous in the presence of diffusion.
\end{abstract}
\keywords{stochastic heat equation, stochastic amplifier, intermittency}
\maketitle
%
%%%%%%%%%%%%%%%%%%%%
%
\section{Introduction}\label{sec0}
Stochastic amplification refers to situations where a fluctuating environment $S(x,t)$ drives the growth of a physical field $\psi(x,t)$. The action of the external random field $S(x,t)$ leads to a net increase of the ``mass'' of $\psi(x,t)$, defined as its spatial average at time $t$. Since $S(x,t)$ is random, the amplification of $\psi(x,t)$ is not uniform and the values of $\psi(x,t)$ at different points contribute differently to the mass. When the mass is dominated entirely by the contribution of the highest peaks of $\psi(x,t)$ that occupy a negligible portion of space, we say that $\psi(x,t)$ is (spatially) intermittent. Intermittency can occur either asymptotically as $t\to\infty$ or at finite time, depending on how $S(x,t)$ acts on $\psi(x,t)$.

Finite time intermittency was proved recently in \cite{Mounaix2026} for the stochastic heat equation,
\begin{equation}\label{withDeq}
\left\lbrace
\begin{array}{l}
\partial_t\psi(x,t)-\frac{1}{2}\partial^2_{x^2} \psi(x,t)=g\, S(x,t)^2\, \psi(x,t), \\
0\le t\le T,\ x\in\mathbb{R},\ {\rm and}\ \psi(x,0)=1,
\end{array}\right.
\end{equation}
at a fixed $T>0$, depending on the coupling constant $g>0$. In equation\ (\ref{withDeq}), amplification is accompanied by diffusion, which disperses the field $\psi$ in space, and $S(x,t)$ is a real-valued Gaussian random field defined by
\begin{equation}\label{defofS}
\begin{array}{l}
\langle S(x,t)\rangle =0, \\
\langle S(x,t)S(x^\prime,t^\prime)\rangle =C(x-x^\prime,t-t^\prime),
\end{array}
\end{equation}
with normalization $C(0,0)=1$. Here and below, $\langle\cdot\rangle$ denotes the average over realizations of $S$. The diffraction version of (\ref{withDeq}) -- obtained by replacing $\partial^2_{x^2}$ with $i\partial^2_{x^2}$ -- which models the backscattering of an incoherent laser by an optically active medium was first considered by Akhmanov et al. in nonlinear optics \cite{ADP1974}, and by Rose and DuBois in laser-plasma interaction \cite{RD1994}. There is a large literature on this topic, and we refer the interested reader to \cite{Mounaix2025} and references therein for background. Considering a diffusive process has the advantage of allowing an easier control of the amplification of $\psi$, something that would be much more challenging to achieve in the diffractive case \cite{ADLM2001,MCL2006}. The stochastic heat equation (\ref{withDeq}) is thus a promising starting point for gaining insight into the more complex diffraction-amplification problem.

For notational convenience, we define $\mathcal{E}(x,g)\equiv \psi(x,T)$, making the dependence on $g$ explicit (and the dependence on $T$ implicit). The analysis of (\ref{withDeq}) in \cite{Mounaix2026} establishes that, for spatially ergodic $S(x,t)$ and fixed $T>0$, there is a critical coupling $g_c>0$, given by
\begin{equation}\label{critcoupling}
g_c=\inf\{\,  g>0\, :\, \langle\mathcal{E}(0,g)\rangle=+\infty\, \} ,
\end{equation}
such that: (i) in the subcritical regime $g<g_c$, $\mathcal{E}(x,g)$ is ergodic and non-intermittent, and (ii) in the supercritical regime $g>g_c$, $\mathcal{E}(x,g)$ is intermittent with loss of ergodicity. In case (i), the high peaks of $\mathcal{E}(x,g)$ are too rare to contribute significantly to the mass
\begin{equation}\label{defofmass}
m(g)=\lim_{L\to\infty}\, \frac{1}{L}\, \int_{-L/2}^{L/2}\mathcal{E}(x,g)\, dx,
\end{equation}
whereas in case (ii), the mass is dominated by the highest peaks of $\mathcal{E}(x,g)$, even though they occupy only a negligible fraction of space. In other words, at fixed $T>0$, there is a transition to intermittency as $g$ increases past $g_c$.

It is worth noting that these results stand in contrast to the asymptotic intermittency phenomenon, which has been extensively studied since the late 1980s for the stochastic heat equation (\ref{withDeq}) with $S(x,t)^2$ replaced by $S(x,t)$; see, e.g., \cite{GM1990,Molchanov1991,Mikhailov1989,Mikhailov1991,BC1995,Noble1997,BG1998} and the references therein. In the latter case, there is no finite-time intermittency for any $g>0$: intermittency emerges only asymptotically as $T\to +\infty$.

To characterize the transition, we define the order parameter
\begin{equation}\label{orderparamnoabs}
Q(g)=\frac{\exp(gT)}{\langle\mathcal{E}(0,g)\rangle},
\end{equation}
which quantifies the uniformity of the spatial distribution of the mass. The numerator corresponds to the amplification obtained when the fluctuations of $S(x,t)^2$ around $\langle S(x,t)^2\rangle =1$ are neglected, while, by the results of \cite{Mounaix2026}, the denominator here is equivalent to the mass. This interchangeability of $\langle\mathcal{E}(0,g)\rangle$ and $m(g)$ in (\ref{orderparamnoabs}) follows from ergodicity in the subcritical regime and from the divergence of both quantities in the supercritical regime. $Q(g)=1$ indicates that the fluctuations of $S^2$ contribute negligibly to the mass, which in this limit is uniformly distributed in space. In contrast, $Q(g)=0$ implies an intermittent $\mathcal{E}(x,g)$ whose mass is highly non-uniformly distributed. Note that $Q(g)\le 1$ by Jensen's inequality.

In order to gain a better understanding of how diffusion can affect the transition to intermittency, it is instructive to first consider the diffusion-free problem,
\begin{equation}\label{withoutDeq}
\left\lbrace
\begin{array}{l}
\partial_t\psi(x,t)=g\, S(x,t)^2\, \psi(x,t), \\
0\le t\le T,\ x\in\mathbb{R},\ {\rm and}\ \psi(x,0)=1,
\end{array}\right.
\end{equation}
which yields
\begin{equation}\label{withoutDsol}
\mathcal{E}(x,g)={\rm e}^{g\int_0^T S(x,\tau)^2\, d\tau}.
\end{equation}
Averaging (\ref{withoutDsol}) over $S$ gives
\begin{equation}\label{withoutDaverage}
\langle\mathcal{E}(0,g)\rangle=\left\lbrace
\begin{array}{cr}
Z(g)\, \left(1-2g\mu_1\right)^{-1/2},&\forall g<1/2\mu_1, \\
+\infty ,&\forall g\ge 1/2\mu_1,
\end{array}\right.
\end{equation}
where $\mu_1>\mu_2\ge\cdots\ge 0$ are the eigenvalues of the kernel $C(0,t-t^\prime)$ on $0\le t,\, t^\prime\le T$, and $Z(g)=\prod_{n\ge 2}(1-2g\mu_n)^{-1/2}$ is bounded on $[0,1/2\mu_1]$. (Note that $\mu_1$ is not degenerate for the class of $S$ we will consider.) Equations (\ref{critcoupling}) and (\ref{withoutDaverage}) yield $g_c=1/2\mu_1$, and using the expression (\ref{withoutDaverage}) in equation (\ref{orderparamnoabs}), one obtains
\begin{equation}\label{QofgwithoutD}
Q(g)=\left\lbrace
\begin{array}{cr}
Z(g)^{-1}\exp(gT)\, \sqrt{(g_c-g)/g_c},&\forall g<g_c, \\
0 ,&\forall g\ge g_c.
\end{array}\right.
\end{equation}
\begin{figure}[!h]
\includegraphics[width = 0.6\linewidth]{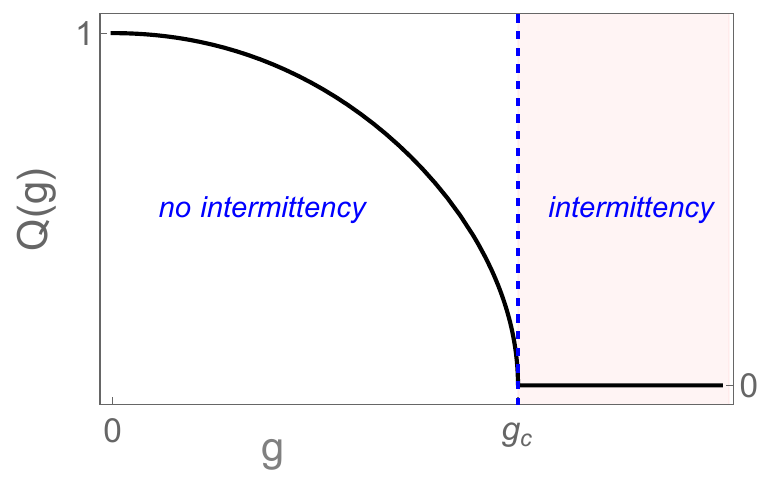}
\caption{Order parameter $Q(g)$ as a function of the control parameter $g$ in the diffusion-free limit. The transition between intermittent and non-intermittent regimes at $g=g_c$ is continuous.}\label{figure1}
\end{figure}

Figure~\ref{figure1} shows the behavior of $Q(g)$ given by equation (\ref{QofgwithoutD}). As $g$ increases, $Q(g)$ decreases continuously from $Q(0)=1$ to $Q(g_c)=0$, and remains zero for all $g\ge g_c$. Therefore, in the diffusion-free limit, the transition between intermittent and non-intermittent regimes is continuous, in the sense that $Q(g)$ is continuous at $g_c$. Since $Q(g_c)=0$, the intermittent regime extends down to and including $g=g_c$.

As $g$ approaches $g_c$ from below, $\mathcal{E}(x,g)$ becomes increasingly non-uniform, eventually indistinguishable from a genuinely intermittent field. We will refer to this characteristic feature of the continuous transition as a ``subcritical precursor'' of intermittency. Interestingly, this scenario closely parallels that of asymptotic intermittency, with the limit $T\to\infty$ replaced by $g\to g_c^{-}$. From this perspective, asymptotic intermittency may itself be interpreted as the subcritical precursor of a fully developed intermittency occurring at $T=\infty$. We now examine how the picture is modified when diffusion is introduced.

In the presence of diffusion, equation (\ref{withoutDsol}) is replaced with the Feynman-Kac formula,
\begin{equation}\label{noabssol}
\mathcal{E}(x,g)=
\int_{B(x,T)} {\rm e}^{g\int_0^T S(x(\tau),\tau)^2\, d\tau}\, d\mathbb{P}_W[x(\cdot)],
\end{equation}
where $B(x,T)$ denotes the set of continuous paths such that $x(T)=x$ and $d\mathbb{P}_W$ is the Wiener measure.  Averaging (\ref{noabssol}) over $S$ yields
\begin{equation}\label{withDaveragenoabs}
\langle\mathcal{E}(0,g)\rangle=\int_{B(0,T)}\mathcal{A}[x(\cdot); g]\, d\mathbb{P}_W[x(\cdot)],
\end{equation}
where
\begin{equation}\label{pathampli}
\mathcal{A}[x(\cdot); g]=\left\lbrace
\begin{array}{cr}
Z[x(\cdot); g]\, \left(1-2g\mu_1[x(\cdot)]\right)^{-1/2},&\forall g<1/2\mu_1[x(\cdot)], \\
+\infty ,&\forall g\ge 1/2\mu_1[x(\cdot)].
\end{array}\right.
\end{equation}
Here, $\mu_1[x(\cdot)]>\mu_2[x(\cdot)]\ge\cdots\ge 0$ are the eigenvalues of the kernel $C(x(t)-x(t^\prime),t-t^\prime)$ on $0\le t,\, t^\prime\le T$, and
$$
Z[x(\cdot); g]=\prod_{n\ge 2}(1-2g\mu_n[x(\cdot)])^{-1/2}
$$
is bounded for $g$ in $[0,1/2\mu_1[x(\cdot)]]$. (Again, $\mu_1[x(\cdot)]$ is not degenerate for the class of $S$ we will consider.)

The first effect of diffusion on the transition is a possible reduction in the critical coupling, which can never exceed its value in the absence of diffusion \cite{ADLM2001}. Indeed, the argument in \cite{Mounaix2026} yields $g_c=1/2\mu_{\rm max}$, where $\mu_{\rm max}=\sup_{x\in B(0,T)}\mu_1[x(\cdot)]$. Since $\mu_{\rm max}\ge\mu_1[x(\cdot)\equiv 0]=\mu_1$, the critical coupling in the presence of diffusion is bounded above by its diffusion-free counterpart, $1/2\mu_1$. This somewhat surprising result arises because diffusion allows the system to explore paths along which the amplification can be larger than for constant paths. By contrast, the diffusion-free problem only probes constant paths and may therefore miss those that maximize the amplification.

\begin{figure}[!h]
\includegraphics[width = 0.6\linewidth]{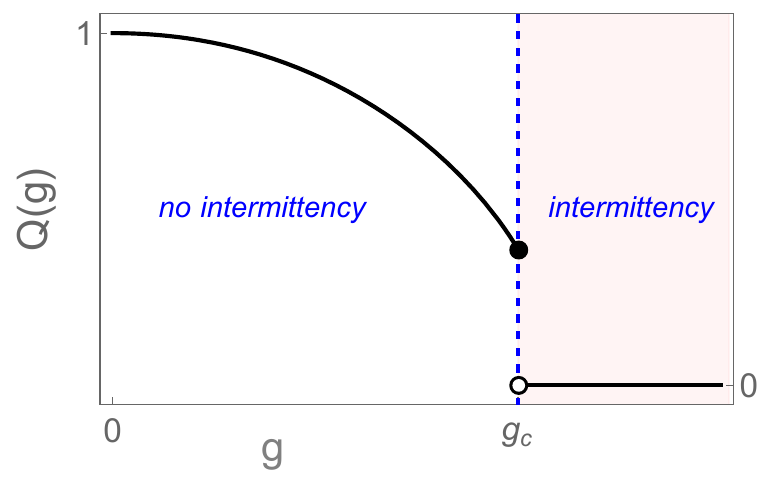}
\caption{Order parameter $Q(g)$ as a function of the control parameter $g$ in the case of a discontinuous transition at $g=g_c$, with $g_c$ not exceeding its diffusion-free counterpart. Non-intermittent regime extends up to and including $g=g_c$ (black end point), which is excluded from the intermittent regime (white point).}\label{figure2}
\end{figure}

The second effect of diffusion concerns the value of $\langle\mathcal{E}(0,g_c)\rangle$, which, according to (\ref{withDaveragenoabs}) and (\ref{pathampli}), is given by
\begin{equation}\label{criticalwithDaveragenoabs}
\langle\mathcal{E}(0,g_c)\rangle=\int_{B(0,T)}
\frac{Z\left[x(\cdot); g_c\right]\, \sqrt{\mu_{\rm max}}}{\sqrt{\mu_{\rm max}-\mu_1[x(\cdot)]}}\, 
d\mathbb{P}_W[x(\cdot)].
\end{equation}
Unlike in the diffusion-free case where $\langle\mathcal{E}(0,g_c)\rangle =\infty$, the presence of diffusion allows for an additional possibility: the singularity at $\mu_1[x(\cdot)]\to\mu_{\rm max}$ in the integrand of (\ref{criticalwithDaveragenoabs}) can be Wiener-integrable. If so, this reflects a strong damping effect of diffusion on the amplification as $g$ approaches $g_c$ from below, resulting in $\langle\mathcal{E}(0,g_c)\rangle <\infty$ and hence $Q(g_c)>0$. The transition is then discontinuous, in the sense that $Q(g)$ is discontinuous at $g_c$, as illustrated in figure~\ref{figure2}. Since $Q(g_c)>0$, the non-intermittent regime extends up to and including $g=g_c$.

Having $Q(g_c)>0$ implies that $\mathcal{E}(x,g)$ in the subcritical regime does not become non-uniform enough to resemble a fully developed intermittent field, even when $g$ is arbitrarily close to $g_c$. In other words, the subcritical precursor is absent in a discontinuous transition: intermittency remains invisible from the subcritical side, up to and including $g=g_c$.

The question is now whether diffusion can indeed make the transition to intermittency discontinuous. The goal of this work is to show that this is actually the case. Addressing this question for equation (\ref{withDeq}), however, leads to difficulties that we are unable to resolve at the present time. As explained in detail in section \ref{sec6}, these difficulties stem from the fact that the Feynman-Kac representation of $\mathcal{E}(x,g)$ involves Brownian paths with arbitrarily large ranges. This is why, in section \ref{sec1a}, we introduce a modified version of (\ref{withDeq}) for which the range of the Brownian paths is naturally bounded. In this regularized setting, the calculation can be carried out in full, and the discontinuity of the transition can be established.

Before going into the details of the calculations, we briefly summarize our main results in qualitative terms. A more detailed summary will be given at the end of section \ref{sec1c}, once all the necessary background have been established. The existence of $\langle\mathcal{E}(0,g_c)\rangle$ in (\ref{criticalwithDaveragenoabs}) depends on the contribution of the Brownian paths for which $Y[x(\cdot)]=\mu_{\rm max}-\mu_1[x(\cdot)]$ is close to zero. In view of this observation, we propose to tackle the problem by estimating the cumulative distribution function (CDF) of the random variable $Y[x(\cdot)]$ in the limit where its argument tends to zero. Along these lines, we first establish that, for small arguments, the CDF of $Y[x(\cdot)]$ can be bounded in terms of a combination of probabilities involving global geometric characteristics of Brownian trajectories. We then apply classical estimates for Brownian motion to these probabilities and show that the CDF of $Y[x(\cdot)]$ decays faster than any power as its argument tends to zero. This rapid decay ensures the convergence of the path integral in (\ref{criticalwithDaveragenoabs}). Consequently, $\langle\mathcal{E}(0,g_c)\rangle <\infty$, from which we conclude that the transition to intermittency is discontinuous.

The paper is organized as follows. In section \ref{sec1}, we defined the modified version of (\ref{withDeq}) (section \ref{sec1a}), we outline our strategy to prove the discontinuity of the transition (section \ref{sec1b}), and we give a more detailed summary of our main results (section \ref{sec1c}). In section \ref{sec2}, we specify the class of $S$ we consider (section \ref{sec2a}) and we determine the corresponding critical coupling $g_c$ (section \ref{sec2b}). Sections \ref{sec3} and \ref{sec4} are devoted to the derivation of the inequalities needed to bound the CDF of $Y[x(\cdot)]$ for small arguments. In section \ref{sec5}, we give the control of the CDF of $Y[x(\cdot)]$ as its argument tends to zero and show that the transition to intermittency is discontinuous. Finally, we discuss our results and outline possible perspectives in section \ref{sec6}. Selected technical material is collected in the appendices.
%
%%%%%%%%%%%%%%%%%%%%
%
\section{Preliminaries}\label{sec1}
%
%%%%%%%%%%
%
\subsection{Setting up the problem}\label{sec1a}
To circumvent the difficulties inherent in problem (\ref{withDeq}), we consider instead a toy model combining two desirable features. First, it retains the full set of results established in \cite{Mounaix2026}. Second, it permits a complete analytical treatment by imposing a natural cut-off on the range of the Brownian paths in the associated Feynman-Kac formula. More specifically, we assume that the diffusion-amplification process (\ref{withDeq}) is embedded in a random lattice of point absorbers, and we consider the random field $\psi(x,t)\equiv \langle\phi(x,t)\rangle_{a.p.}$, where $\langle\cdot\rangle_{a.p.}$ denotes the average over absorber positions and $\phi(x,t)$ is the solution to
\begin{equation}\label{modifiedwithDeq}
\left\lbrace
\begin{array}{l}
\partial_t\phi(x,t)-\frac{1}{2}\partial^2_{x^2} \phi(x,t)=\left[g\, S(x,t)^2-V(x)\right]\, \phi(x,t), \\
0\le t\le T,\ x\in\mathbb{R},\ {\rm and}\ \phi(x,0)=1.
\end{array}\right.
\end{equation}
In equation\ (\ref{modifiedwithDeq}), $V(x)$ is the absorbing potential
\begin{equation}\label{abspotential}
V(x)=\sum_{n\in\mathbb{Z}} \lim_{\varepsilon\to 0}U_{\varepsilon}(x-x_n),
\end{equation}
with
\begin{equation}\label{potentialU}
U_{\varepsilon}(x)=\left\lbrace
\begin{array}{ll}
0\, &{\rm if}\ |x|>\varepsilon/2, \\
\infty\, &{\rm if}\ |x|\le\varepsilon/2.
\end{array}\right.
\end{equation}
The random positions of the absorbers, denoted by $x_n$ in (\ref{abspotential}), are such that the joint distribution of the spacings $d_n\equiv x_n-x_{n-1}>0$ is invariant under translations of the index. Thus, the $d_n$ are identically distributed random variables, although not necessarily independent. We assume that their common distribution is given by a compactly supported probability measure $\pi$ and we define
\begin{equation}\label{upperboundsupport}
r=\inf\lbrace \chi >0\, :\, \pi(d>\chi)=0\rbrace ,
\end{equation}
the upper bound of its support. Finally, conditional on $d_0$, $x_0$ is taken to be uniformly distributed on $[0,d_0)$. Note that since $0\le x_0 <d_0$, $x_0$ is almost surely the position of the first absorber to the right of $x=0$. The main effect of the presence of the absorbers is to restrict the Feynman-Kac representation of $\mathcal{E}(x,g)$ to Brownian paths whose range does not exceed $r$ (see equations (\ref{quenchedamplisol}) and (\ref{latticeaverage1})), as desired.

The same analysis as in \cite{Mounaix2026} can then be carried out, mutatis mutandis, leading to the same conclusions with $g_c$ given by (\ref{critcoupling}) and order parameter
\begin{equation}\label{orderparameter}
Q(g)=\frac{\mathcal{E}(0,0)\, \exp(gT)}{\langle\mathcal{E}(0,g)\rangle}.
\end{equation}
In this setting, the counterpart of (\ref{noabssol}) reads
\begin{equation}\label{quenchedamplisol}
\mathcal{E}(x,g)=
\int_{B(x,T)} {\rm e}^{g\int_0^T S(x(\tau),\tau)^2\, d\tau}
\, \left\langle {\rm e}^{ -\int_0^T V(x(\tau))\, d\tau}\right\rangle_{a.p.}\, d\mathbb{P}_W[x(\cdot)].
\end{equation}
Averaging (\ref{quenchedamplisol}) over $S$, we obtain
\begin{equation}\label{withDaverage}
\langle\mathcal{E}(0,g)\rangle=\int_{B(0,T)}\mathcal{A}[x(\cdot); g]\, 
\left\langle {\rm e}^{ -\int_0^T V(x(\tau))\, d\tau}\right\rangle_{a.p.}\, d\mathbb{P}_W[x(\cdot)],
\end{equation}
with $\mathcal{A}[x(\cdot); g]$ given by (\ref{pathampli}). In particular, for $g=g_c=1/2\mu_{\rm max}$, equation (\ref{withDaverage}) gives
\begin{equation}\label{criticalwithDaverage}
\langle\mathcal{E}(0,g_c)\rangle=\int_{B(0,T)}
\left\langle {\rm e}^{ -\int_0^T V(x(\tau))\, d\tau}\right\rangle_{a.p.}\, 
\frac{Z\left[x(\cdot); g_c\right]\, \sqrt{\mu_{\rm max}}}{\sqrt{\mu_{\rm max}-\mu_1[x(\cdot)]}}\, 
d\mathbb{P}_W[x(\cdot)].
\end{equation}
%
%%%%%%%%%%
%
\subsection{Plan of attack}\label{sec1b}
In light of the discussion in section \ref{sec0}, to establish the discontinuity of the transition we need to show that $\langle\mathcal{E}(0,g_c)\rangle<\infty$. One possible strategy is as follows. Let us denote by $R_T[x(\cdot)]=\sup_{0\le t\le T}x(t)-\inf_{0\le t\le T}x(t)$ the range of $x(\cdot)$ and by $B_r(0,T)$ the set of all paths in $B(0,T)$ satisfying $R_T[x(\cdot)] \le r$. Write $\mu^\prime_{\rm max}=\sup_{x\in B_r(0,T)}\mu_2[x(\cdot)]$. We first obtain an upper bound on $\langle\mathcal{E}(0,g_c)\rangle$ by successively applying the estimates
\begin{equation}\label{boundingabsav}
\left\langle {\rm e}^{ -\int_0^T V(x(\tau))\, d\tau}\right\rangle_{a.p.}\le
\bm{1}_{R_T[x(\cdot)]\le r},
\end{equation}
then
\begin{equation}\label{boundingZ}
\bm{1}_{R_T[x(\cdot)]\le r}\, Z\left[x(\cdot); g_c\right]\le
\exp\left(\frac{g_cT}{1-\mu^\prime_{\rm max}/\mu_{\rm max}}\right) ,
\end{equation}
to the corresponding terms on the right-hand side of (\ref{criticalwithDaverage}). The bound in (\ref{boundingabsav}) is a key element of our proof; its derivation is given in appendix \ref{app1}. The one in (\ref{boundingZ}) follows from $-\ln(1-x)\le x/(1-x)$ and $\sum_{n=1}^{+\infty}\mu_n[x(\cdot)]=C(0,0)\, T$, with normalization $C(0,0)=1$. We thus obtain
\begin{equation}\label{boundingE}
\langle\mathcal{E}(0,g_c)\rangle\le
\sqrt{\mu_{\rm max}}\, \exp\left(\frac{g_cT}{1-\mu^\prime_{\rm max}/\mu_{\rm max}}\right)\, 
\int_{B(0,T)}\frac{\bm{1}_{R_T[x(\cdot)]\le r}\, d\mathbb{P}_W[x(\cdot)]}{\sqrt{\mu_{\rm max}-\mu_1[x(\cdot)]}}.
\end{equation}
For the class of $S$ we will consider, $\mu_1[x(\cdot)]$ is not degenerate and for all $x(\cdot)$ in $B_r(0,T)$, one has the inequality,
\begin{equation}\label{spectralgap}
\frac{\mu_2[x(\cdot)]}{\mu_1[x(\cdot)]}\le 1-C(r,T),
\end{equation}
the proof of which is given in appendix \ref{appSG}. From (\ref{spectralgap}) it follows that
$$
\frac{\mu_2[x(\cdot)]}{\mu_{\rm max}}\le 1-C(r,T),
$$
and taking the supremum over $B_r(0,T)$ yields $\mu^\prime_{\rm max}/\mu_{\rm max}\le 1-C(r,T)<1$. This ensures that the prefactor in (\ref{boundingE}) is finite and we are done if we can prove that
\begin{equation}\label{firstboundedintegral}
\int_{B(0,T)}\frac{\bm{1}_{R_T[x(\cdot)]\le r}\, d\mathbb{P}_W[x(\cdot)]}{\sqrt{\mu_{\rm max}-\mu_1[x(\cdot)]}}<+\infty .
\end{equation}
It is convenient to rewrite (\ref{firstboundedintegral}) in a more tractable form. We define
\begin{equation}\label{definitionofYofx}
Y[x(\cdot)]=\mu_{\rm max}-\mu_1[x(\cdot)],
\end{equation}
and we write
\begin{equation}\label{regularizedCDFY}
F(y)=P\left(Y[x(\cdot)]< y\ {\rm and}\ R_T[x(\cdot)]\le r\right),
\end{equation}
where
\begin{equation}\label{regularizedCDFdefinition}
P\left(Y[x(\cdot)]< y\ {\rm and}\ R_T[x(\cdot)]\le r\right)=
\int_{B(0,T)}\bm{1}_{Y[x(\cdot)]< y}\, \bm{1}_{R_T[x(\cdot)]\le r}\, d\mathbb{P}_W[x(\cdot)].
\end{equation}
In terms of $Y[x(\cdot)]$, equation (\ref{firstboundedintegral}) takes the compact form
\begin{equation}\label{secondboundedintegral}
\int_0^{\mu_{\rm max}}\frac{dF(y)}{\sqrt{y}}<+\infty ,
\end{equation}
where the domain of integration follows from $0<\mu_1[x(\cdot)]\le\mu_{\rm max}$. Since for any $\varepsilon >0$,
$$
\int_\varepsilon^{\mu_{\rm max}}\frac{dF(y)}{\sqrt{y}}\le\frac{1}{\sqrt{\varepsilon}}\, \int_\varepsilon^{\mu_{\rm max}}dF(y)\le\frac{1}{\sqrt{\varepsilon}} <\infty,
$$
the convergence of the integral in (\ref{secondboundedintegral}) depends on the integrability near $y=0$, i.e., on the behavior of $F(y)$ for small values of $y$. The next step is thus clear: estimate $F(y)$ near $y=0$ and show that it vanishes fast enough as $y\to 0$ for the integral to converge.

\begin{figure}[!h]
\includegraphics[width = 0.65\linewidth]{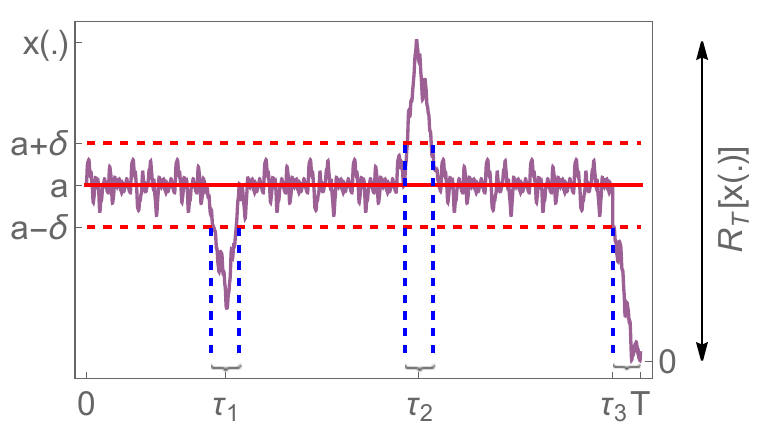}
\caption{Schematic representation of a Brownian path $x(\cdot)$ in $B(0,T)$ (i.e., such that $x(T)=0$) staying within the distance $\delta$ of a constant position $a$ for time $T-\tau$, with $\tau=\sum_i \tau_i$, where the $\tau_i$ are the durations of the time intervals where $x(t)$ deviates from $a$ by more than $\delta$. For the case considered in the text, $Y[x(\cdot)]\to 0$ as $\delta,\, \tau\to 0$, even though the range $R_T[x(\cdot)]$ (indicated by a double arrow on the right of the figure) remains finite: $Y[x(\cdot)]\to 0$ does {\it not} imply $R_T[x(\cdot)]\to 0$.}\label{figure3}
\end{figure}

To carry out this strategy, we need to identify the paths for which $Y[x(\cdot)]$ is small. Suppose, for the sake of discussion, that $\mu_1[x(\cdot)]$ attains its maximum on constant paths. Intuitively, it is clear that a small value of $Y[x(\cdot)]$ corresponds to $x(\cdot)$ ``close'' to a constant path in a sense to be specified. It was proved in \cite{Mounaix2026} that $\mu_1[x(\cdot)]$ is continuous on $B(0,T)$ endowed with the uniform norm. Thus, a small range $R_T[x(\cdot)]$ implies a small $Y[x(\cdot)]$, but the converse is not true. To see why, consider a continuous path that stays within the distance $\delta$ of a constant $a$ for time $T-\tau$, and deviates from $a$ by more than $\delta$ only during brief intervals of total duration $\tau$ (see figure \ref{figure3} for a pictorial representation of these quantities). By choosing $\tau$ and $\delta$ sufficiently small, $\mu_1[x(\cdot)]$ can be made arbitrarily close to $\mu_{\rm max}$. The limiting path is not continuous, but $Y[x(\cdot)]=0$ can be approached arbitrarily closely by continuous paths for which $R_T[x(\cdot)]$ does not vanish. Such a possibility must be taken into account in the estimation of $F(y)$ near $y=0$. Loosely speaking, a small $Y[x(\cdot)]$ does not require $x(\cdot)$ to remain uniformly close to a constant. Short excursions of finite amplitude are still possible. It remains to formalize this idea and make it operational. This is done in the present paper for $S$ in a wide class of physical interest.
%
%%%%%%%%%%
%
\subsection{Detailed summary of the main results}\label{sec1c}
We are now in a position to provide a more detailed and technical summary of our main results, complementing the qualitative overview presented at the end of section \ref{sec0}. For the class of $S$ specified in section \ref{sec2a}, we first establish that, for $y$ small enough,
\begin{equation}\label{boundingFintro}
F(y)\le P\bigl[\, \nu_{x(\cdot)}(c(r)\, y^{1/4})<\sqrt{y}\, \bigr],
\end{equation}
where $c(r)$ is defined in (\ref{constantcofr}), and
\begin{equation}\label{defofnu}
\nu_{x(\cdot)}(\delta)=\min_{0\le s\le T}\int_0^T\bm{1}_{|x(s)-x(u)|>\delta}\, du,
\end{equation}
is the minimal $\delta$-excursion measure of $x(\cdot)\in B(0,T)$. The probability on the right-hand side of (\ref{boundingFintro}) does not contain the joint condition $R_T[x(\cdot)]\le r$, unlike $F(y)$ on the left-hand side (see equation (\ref{regularizedCDFY})). So, in the expression $\nu_{x(\cdot)}(c(r)\, y^{1/4})$ on the right-hand side of (\ref{boundingFintro}), the Brownian path $x(\cdot)$ is not subject to any restriction on its range, and $r$ is a free parameter. We then show that for any $K>0$,
\begin{equation}\label{boundprobanuintro}
P\left(\nu_{x(\cdot)}(\delta)<\tau\right)\le P\left(\|x(\cdot)\|_\infty\le2(\delta + K)\right)
+P\left(\omega_{x(\cdot)}(\tau)>K\right) ,
\end{equation}
where $\|x(\cdot)\|_\infty =\sup_{0\le t\le T}|x(t)|$ and
\begin{equation}\label{defofomega}
\omega_{x(\cdot)}(\tau) = \sup_{\substack{0\le s,u\le T\\ \vert s-u\vert\le\tau}}\vert x(s)-x(u)\vert ,
\end{equation}
is the modulus of continuity of $x(\cdot)$ at scale $\tau$. Finally, we use (\ref{boundprobanuintro}) with $K\propto y^{1/8}$ to bound the right-hand side of (\ref{boundingFintro}) and, by applying classical estimates for Brownian motion, we obtain the control of $F(y)$ near $y=0$ (equation (\ref{finalcontrolofF})),
$$
F(y)\le\frac{4}{\pi}\, \left(1+6\sqrt{\pi}\, \frac{(2T/3)^{3/4}}{y^{3/8}}\right)
\, \exp\left(-\frac{\sqrt{2T/3}}{16\, y^{1/4}}\right) ,
$$
which goes to zero faster than any power of $y$ as $y\to 0$. This rapid decay ensures the convergence of the integral in (\ref{secondboundedintegral}). Consequently, $\langle\mathcal{E}(0,g_c)\rangle <\infty$, and the transition to intermittency in problem (\ref{modifiedwithDeq}) is discontinuous for the considered class of $S$.
%
%%%%%%%%%%%%%%%%%%%%
%
\section{Specification of $\bm{S(x,t)}$ and determination of the critical coupling}\label{sec2}
%
%%%%%%%%%%
%
\subsection{Specification of $\bm{S(x,t)}$}\label{sec2a}
To proceed without getting unnecessarily lost in the technical intricacies that a fully general treatment of the problem would inevitably entail, we restrict ourselves to a particular class of $S$. This class captures the essential physics at play while providing a good compromise between keeping the calculations as simple as possible and retaining physically realistic features.

Scale-mixture representations of correlation functions -- such as the Mat\'ern family and its generalizations -- are powerful tools for simulating Gaussian fields with flexible correlation structures (see, e.g., \cite{BECP2025,Matern1986} and references therein). Drawing loosely on these models, we consider a Gaussian field $S(x,t)$ whose correlation function is a mixture of separable correlations,
\begin{equation}\label{mixturecorrel1}
C(x,t)=\int_{\kappa,\, \omega\, >\, 0} \mathscr{C}(\kappa x)\, \mathscr{T}(\omega t)\, d\rho(\kappa,\, \omega),
\end{equation}
where $\mathscr{C}(x)$ and $\mathscr{T}(t)$ are correlation functions, and $d\rho(\kappa,\, \omega)$ is a joint probability measure of $\kappa>0$ and $\omega>0$.

In the particular case where $d\rho(\kappa,\, \omega)=\delta(\omega -\alpha(\kappa))\, d\rho(\kappa)\, d\omega$ with $\alpha(\kappa)>0$, this expression reduces to
\begin{equation}\label{mixturecorrel2}
C(x,t)=\int_0^{\infty} \mathscr{C}(\kappa x)\, \mathscr{T}(\alpha(\kappa)t)\, d\rho(\kappa),
\end{equation}
which can be viewed as a real-space counterpart of spectral superpositions of modes with scale-dependent time dynamics. In turbulence modeling, for instance, spectral representations of velocity fields involve such superpositions, where eddies of size $\kappa^{-1}$ persist for a scale-dependent time $\alpha(\kappa)^{-1}$, as in Kraichnan-type models  \cite{Kraichnan1964,Kraichnan1970} or random sweeping approaches \cite{Tennekes1975}. In physical space, this leads to correlation functions that can be expressed as mixtures of separable space–time contributions. Similar constructions also appear in the theory of correlation functions for space-time data (see, e.g., \cite{Gneiting2002}).

The expression in (\ref{mixturecorrel1}) generalizes (\ref{mixturecorrel2}) by allowing spatial and temporal scales to be jointly distributed, rather than being constrained by $\omega=\alpha(\kappa)$. The flexibility in the choice of $\mathscr{C}$, $\mathscr{T}$, and $d\rho$ makes it versatile enough for modeling correlations in a wide range of physical situations.

For our purposes, we impose conditions on $\mathscr{C}$ and $\mathscr{T}$, while $d\rho$ is left unrestricted. The first two conditions are required for the results of \cite{Mounaix2026} to hold, and are equivalent to the spectral density conditions assumed in section 2.1 of that work.
\begin{condition}\label{condition1}
(Ergodicity). $\mathscr{C}(x)$ goes to zero as $|x|\to\infty$.
\end{condition}
\begin{condition}\label{condition2}
(Smoothness and normalization). $\mathscr{C}(x)$ is an even $C^4(\mathbb{R})$ function normalized to $\mathscr{C}(0)=1$, and  $\mathscr{T}(t)$ is an even $C^4([-T,\, T])$ function normalized to $\mathscr{T}(0)=1$.
\end{condition}
\noindent N.B.: as correlation functions, $\mathscr{C}$ and $\mathscr{T}$ satisfy $|\mathscr{C}(x)|\le\mathscr{C}(0)$ and $|\mathscr{T}(t)|\le\mathscr{T}(0)$.

\bigskip
Condition~\ref{condition1} is equivalent to the Gaussian field $S$ being ergodic with respect to spatial translations \cite{Maruyama1949,Grenander1950,BE1972}. The smoothness conditions~\ref{condition2} ensure that $S\in C^2(\mathbb{R}^2)$ almost surely \cite{Adler1981,Delmas1998}. The normalizations of $\mathscr{C}$ and $\mathscr{T}$ in condition~\ref{condition2}, together with $\int d\rho(\kappa,\, \omega)=1$, ensure $C(0,0)=1$. The specification of $S$ is completed by the following additional conditions.
\begin{condition}\label{condition3}
$\mathscr{C}(x)$ is positive and decreasing for $x>0$, and $\mathscr{T}(t)$ is positive, decreasing for $t>0$, and uniformly bounded away from zero on $0\le t\le T$.
\end{condition}
\begin{condition}\label{condition4}
$\partial_{x^2}^2\mathscr{C}(0)$ is negative and bounded away from zero.
\end{condition}
%
%%%%%%%%%%
%
\subsection{Determination of the critical coupling}\label{sec2b}
As a first, preliminary result, we now show that for this class of $S$, the value of the critical coupling in the presence of diffusion is equal to the one in the diffusion-free case. From conditions \ref{condition2} and \ref{condition3} it follows that, for any $x(\cdot)$ in $B(0,T)$, $C(x(t)-x(t^\prime),t-t^\prime)$ is a positive, compact kernel on $0\le t,\, t^\prime\le T$ whose largest eigenvalue is non-degenerate, with a positive eigenfunction (Jentzsch theorem \cite{Jentzsch1912,AA2002}). Write $\varphi(t)$ this eigenfunction normalized to $\|\varphi\|_2=1$, where $\|\cdot\|_2$ denotes the $L^2$-norm on $[0,\, T]$. Using the inequality $C(x(t)-x(t^\prime),t-t^\prime)\le C(0,t-t^\prime)$, which follows from $\mathscr{C}(x)\le\mathscr{C}(0)$ and $\mathscr{T}(t)>0$ (condition \ref{condition3}), and writing $\mu_1$ for $\mu_1[x(\cdot)\equiv 0]$, one has
\begin{eqnarray}\label{mumaxismu1}
\mu_1[x(\cdot)]&=&\int_0^T \int_0^T C(x(t)-x(t^\prime),t-t^\prime)\, \varphi(t)\, \varphi(t^\prime)\, dt\, dt^\prime \nonumber \\
&\le&\int_0^T \int_0^T C(0,t-t^\prime)\, \varphi(t)\, \varphi(t^\prime)\, dt\, dt^\prime \le\mu_1,
\end{eqnarray}
where the last inequality follows from the min--max principle \cite{CH1989,RS1978}. Taking the supremum over $x(\cdot)$ in $B(0,T)$ then yields $\mu_{\rm max}\le\mu_1$, and therefore $\mu_{\rm max}=\mu_1$. Hence, the critical coupling with diffusion, $g_c=1/2\mu_{\rm max}$, is equal to its diffusion-free counterpart, $g_c=1/2\mu_1$, as announced.

This first result shows that for the considered class of $S$, the constant path $x(\cdot)\equiv 0$ maximizes $\mu_1[x(\cdot)]$. Assume now that $x(\cdot)\not\equiv 0$ in equation (\ref{mumaxismu1}). Then the first inequality is in fact strict, as a consequence of $\mathscr{C}(x)<\mathscr{C}(0)$ for $x\neq 0$ (condition \ref{condition3}). This yields $\mu_1[x(\cdot)]<\mu_{\rm max}$, and $x(\cdot)\equiv 0$ is the only path in $B(0,T)$ that maximizes $\mu_1[x(\cdot)]$. We now move on to carrying out the program outlined in the introduction.
%
%%%%%%%%%%
%
\section{From small $\bm{Y[x(\cdot)]}$ to small $\bm{\nu_{x(\cdot)}(\delta)}$}\label{sec3}
In this section, we establish inequality (\ref{boundingFintro}) for $y$ small enough. By the definition of $Y[x(\cdot)]$ and the equality $\mu_{\rm max}=\mu_1$ proved in section \ref{sec2b}, we have, using the same notation as in equation (\ref{mumaxismu1}),
\begin{eqnarray}\label{boundoneY}
&&Y[x(\cdot)]\ge\int_0^T \int_0^T\left[C(0,s-u)-C(x(s)-x(u),s-u)\right]\, \varphi(s)\, \varphi(u)\, ds\, du \nonumber \\
&&\ge\int_0^T \int_0^T\left[C(0,s-u)-C(x(s)-x(u),s-u)\right]\, \bm{1}_{|x(s)-x(u)|>\delta}\, \varphi(s)\, \varphi(u)\, ds\, du,
\end{eqnarray}
where the second inequality follows from the non-negativity of the integrand (by condition~\ref{condition3} and $\varphi(t)>0$). For $|x(s)-x(u)|>\delta$, condition \ref{condition3} implies
\begin{eqnarray*}
C(0,s-u)-C(x(s)-x(u),s-u)&\ge& C(0,s-u)-C(\delta,s-u) \\
&\ge& C(0,T)-C(\delta,T),
\end{eqnarray*}
and using this inequality on the right-hand side of (\ref{boundoneY}) yields
\begin{equation}\label{boundtwoY}
Y[x(\cdot)]\ge\left[C(0,T)-C(\delta,T)\right]
\, \int_0^T \left(\int_0^T \bm{1}_{|x(s)-x(u)|>\delta}\, \varphi(u)\, du\right)\, \varphi(s)\, ds.
\end{equation}
We now derive a lower bound for the integral in (\ref{boundtwoY}). For every $u$ in $[0,T]$ and $x(\cdot)$ in $B_r(0,T)$ (see the beginning of section \ref{sec1b} for the definition of $B_r(0,T)$), condition \ref{condition3} implies
\begin{eqnarray}\label{prelimpointboundofphi}
\varphi(u)&=&\frac{1}{\mu_1[x(\cdot)]}\, \int_0^T C(x(u)-x(s),u-s)\, \varphi(s)\, ds \nonumber \\
&\ge&\frac{C(r,T)}{\mu_1[x(\cdot)]}\, \int_0^T\varphi(s)\, ds.
\end{eqnarray}
Since $C(x,t)\le 1$ and $\varphi(t)>0$, one has
\begin{eqnarray*}
\mu_1[x(\cdot)]&=&\int_0^T \int_0^T C(x(s)-x(u),s-u)\, \varphi(s)\, \varphi(u)\, ds\, du \\
&\le&\left(\int_0^T\varphi(s)\, ds\right)^2,
\end{eqnarray*}
and therefore,
\begin{equation}\label{L1boundofphi}
\int_0^T\varphi(s)\, ds\ge\mu_1[x(\cdot)]^{1/2}.
\end{equation}
Substituting (\ref{L1boundofphi}) into (\ref{prelimpointboundofphi}) gives
\begin{equation}\label{pointboundofphi}
\varphi(u)\ge\frac{C(r,T)}{\mu_1[x(\cdot)]^{1/2}}.
\end{equation}
Combining (\ref{L1boundofphi}), (\ref{pointboundofphi}), and introducing the minimal $\delta$-excursion measure defined in (\ref{defofnu}), one obtains
\begin{equation}\label{boundofintegral}
\int_0^T \left(\int_0^T \bm{1}_{|x(s)-x(u)|>\delta}\, \varphi(u)\, du\right)\, \varphi(s)\, ds\ge C(r,T)\, \nu_{x(\cdot)}(\delta).
\end{equation}
Equations (\ref{boundtwoY}) and (\ref{boundofintegral}) then yield
\begin{equation}\label{boundthreeY}
Y[x(\cdot)]\ge M_r(\delta)\, \nu_{x(\cdot)}(\delta),
\end{equation}
where
\begin{equation}\label{defofM}
M_r(\delta)=C(r,T)\, \left[C(0,T)-C(\delta,T)\right].
\end{equation}
It follows that, for every $x(\cdot)$ in $B_r(0,T)$, $Y[x(\cdot)]<y$ implies $\nu_{x(\cdot)}(\delta)<y/M_r(\delta)$ independent of $x(\cdot)$. We now take advantage of the freedom we have in the choice of $\delta$ to bound $M_r(\delta)$ below by $\sqrt{y}$, when $y$ is sufficiently small.

According to the smoothness condition \ref{condition2} for $\mathscr{C}(x)$, one has the expansion
\begin{equation}\label{taylorexpansion}
C(\delta ,T)=C(0,T)-\frac{1}{2}\, \left|\partial^2_{x^2}C(0,T)\right|\, \delta^2 +o\left(\delta^2\right)
\qquad (\delta\to 0).
\end{equation}
We define
\begin{equation}\label{constantcofr}
c(r)=\frac{2}{\sqrt{C(r,T)\, \left\vert\partial_{x^2}^2 C(0,T)\right\vert}},
\end{equation}
and for a given $y>0$, we take $\delta =c(r)\, y^{1/4}$. From the expression of $M_r(\delta)$ in (\ref{defofM}) and the expansion (\ref{taylorexpansion}) it follows that
$$
M_r\left(c(r)\, y^{1/4}\right)\sim 2\sqrt{y} \qquad (y\to 0),
$$
which means that for every $\alpha >0$, there exists $y_1>0$ such that for all $y\le y_1$,
$$
\left\vert\frac{M_r\left(c(r)\, y^{1/4}\right)}{2\sqrt{y}}-1\right\vert\le\alpha .
$$
In particular,
$$
M_r\left(c(r)\, y^{1/4}\right)\ge (1-\alpha)\, 2\sqrt{y} ,
$$
and taking $\alpha = 1/2$, one obtains
$$
M_r\left(c(r)\, y^{1/4}\right)\ge \sqrt{y} .
$$
Using this result in (\ref{boundthreeY}) one finds that, for every $x(\cdot)$ in $B_r(0,T)$ and $y\le y_1$, $Y[x(\cdot)]<y$ implies $\nu_{x(\cdot)}\left(c(r)\, y^{1/4}\right)<\sqrt{y}$. Hence,
\begin{equation}\label{firstboundingF}
F(y)\le P\bigl[\, \nu_{x(\cdot)}(c(r)\, y^{1/4})<\sqrt{y}\ {\rm and}\ R_T[x(\cdot)]\le r\, \bigr] .
\end{equation}
It remains to bound the right-hand side of (\ref{firstboundingF}) by
$$
P\bigl[\, \nu_{x(\cdot)}(c(r)\, y^{1/4})<\sqrt{y}\ {\rm and}\ R_T[x(\cdot)]\le r\, \bigr]
\le P\bigl[\, \nu_{x(\cdot)}(c(r)\, y^{1/4})<\sqrt{y}\, \bigr]
$$
to obtain
\begin{equation}\label{secboundingF}
F(y)\le P\bigl[\, \nu_{x(\cdot)}(c(r)\, y^{1/4})<\sqrt{y}\, \bigr] ,
\end{equation}
which is the inequality announced in equation (\ref{boundingFintro}).
%
%%%%%%%%%%
%
\section{Intermediate control of $\bm{P\left(\nu_{x(\cdot)}(\delta)<\tau\right)}$}\label{sec4}
In this section, we prove the inequality (\ref{boundprobanuintro}). For notational convenience, we will write $R_T$ for $R_T[x(\cdot)]$. Fix $\delta ,\, \tau>0$, with $\tau<T$, and consider a path $x(\cdot)$ in $B(0,T)$ for which $\nu_{x(\cdot)}(\delta)<\tau$. By the definition of $\nu_{x(\cdot)}(\delta)$, there is an interval of width $2\delta$ inside which $x(\cdot)$ spends at least time $T-\tau$. There are two possibilities: either $x(\cdot)$ never leaves the interval, which is equivalent to $R_T\le 2\delta$, or $x(\cdot)$ leaves the interval, but for no longer than $\tau$, in which case $R_T> 2\delta$. Consequently,
\begin{equation}\label{eq1sec4}
P\left( \nu_{x(\cdot)}(\delta)<\tau\right)=P\left( R_T\le 2\delta\right)
+P\left( \nu_{x(\cdot)}(\delta)<\tau\ {\rm and}\ R_T>2\delta\right).
\end{equation}
We now split the second term on the right-hand side of (\ref{eq1sec4}) according to whether $R_T$ exceeds $2(\delta+K)$, for an arbitrary $K>0$:
\begin{eqnarray}\label{splittingtrick1}
P\left( \nu_{x(\cdot)}(\delta)<\tau\ {\rm and}\ R_T>2\delta\right)&=&
P\left( \nu_{x(\cdot)}(\delta)<\tau\ {\rm and}\ 2\delta<R_T\le 2(\delta+K)\right) \nonumber \\
&+&P\left( \nu_{x(\cdot)}(\delta)<\tau\ {\rm and}\ R_T>2(\delta+K)\right) \nonumber \\
&\le&P\left(2\delta<R_T\le 2(\delta+K)\right) \nonumber \\
&+&P\left( \nu_{x(\cdot)}(\delta)<\tau\ {\rm and}\ R_T>2(\delta+K)\right).
\end{eqnarray}
Combining (\ref{splittingtrick1}) with (\ref{eq1sec4}), we obtain
\begin{eqnarray}\label{sixthbound1}
&&P\left( \nu_{x(\cdot)}(\delta)<\tau\right)\le P\left( R_T\le 2(\delta+K)\right)
+P\left( \nu_{x(\cdot)}(\delta)<\tau\ {\rm and}\ R_T>2(\delta+K)\right) \nonumber \\
&&\le P\left(\| x(\cdot)\|_\infty \le 2(\delta+K)\right)
+P\left( \nu_{x(\cdot)}(\delta)<\tau\ {\rm and}\ R_T>2(\delta+K)\right).
\end{eqnarray}
In the last inequality, we have used $\| x(\cdot)\|_\infty \le R_T$, which follows from the endpoint condition $x(T)=0$. It remains to bound the joint probability $P\left( \nu_{x(\cdot)}(\delta)<\tau\ {\rm and}\ R_T>2(\delta+K)\right)$ by the single-event probability $P\left(\omega_{x(\cdot)}(\tau)>K\right)$.

Let $x(\cdot)$ be a path in $B(0,T)$ such that $R_T>2(\delta+K)$ and $\nu_{x(\cdot)}(\delta)<\tau$. By the definition of $\nu_{x(\cdot)}(\delta)$ in (\ref{defofnu}), there exists a time $s$ in $[0,T]$ such that the set
$$
A=\lbrace 0\le u\le T\, :\, \vert x(u)-x(s)\vert >\delta\rbrace
$$
has length $\vert A\vert <\tau$. Since $R_T>2(\delta+K)$, the point $x(s)$ must be more than $\delta+K$ away from either the maximum or the minimum of the path. By continuity, there is therefore a time $t$ in $[0,T]$ such that
\begin{equation}\label{premierecart}
\vert x(t)-x(s)\vert >\delta+K.
\end{equation}
We now claim that there exists a time $u$ in $[0,T]$ with $\vert t-u\vert\le\tau$ such that
\begin{equation}\label{deuxiemeecart}
\vert x(u)-x(s)\vert\le\delta.
\end{equation}
Suppose that no such $u$ exists. Then every $u$ within $\tau$ of $t$ satisfies $\vert x(u)-x(s)\vert >\delta$, so that
$$
[t-\tau , t+\tau]\cap [0,T] \subseteq A.
$$
Because $\tau<T$, this interval has length at least $\tau$, which implies $\vert A\vert\ge\tau$. This contradicts $\vert A\vert <\tau$. Hence, such a time $u$ must exist. Combining equations (\ref{premierecart}) and (\ref{deuxiemeecart}), we obtain
\begin{eqnarray}\label{troisiemeecart}
\vert x(t)-x(u)\vert&\ge&\vert x(t)-x(s)\vert -\vert x(u)-x(s)\vert \nonumber \\
&>&(\delta+K)-\delta =K.
\end{eqnarray}
Since $\vert t-u\vert\le\tau$, the definition of $\omega_{x(\cdot)}(\tau)$ in (\ref{defofomega}) gives $\omega_{x(\cdot)}(\tau)\ge\vert x(t)-x(u)\vert$, and (\ref{troisiemeecart}) implies $\omega_{x(\cdot)}(\tau)>K$. We have therefore shown that every Brownian path in $B(0,T)$ satisfying $R_T>2(\delta+K)$ and $\nu_{x(\cdot)}(\delta)<\tau$ also satisfies $\omega_{x(\cdot)}(\tau)>K$. Consequently,
\begin{equation}\label{jointprobabound}
P\left( \nu_{x(\cdot)}(\delta)<\tau\ {\rm and}\ R_T>2(\delta+K)\right)\le
P\left(\omega_{x(\cdot)}(\tau)>K\right).
\end{equation}
Finally, using (\ref{jointprobabound}) on the right-hand side of (\ref{sixthbound1}) gives
\begin{equation}\label{ninthbound}
P\left( \nu_{x(\cdot)}(\delta)<\tau \right)\le
P\left(\|x(\cdot)\|_\infty\le 2(\delta +K)\right)
+P\left(\omega_{x(\cdot)}(\tau)>K\right) ,
 \end{equation}
which is the inequality announced in equation (\ref{boundprobanuintro}).
%
%%%%%%%%%%
%
\section{Control of $\bm{F(y)}$ near $\bm{y=0}$ and discontinuous transition}\label{sec5}
In this section, we use the results from sections \ref{sec3} and \ref{sec4} to control the behavior of $F(y)$ near $y=0$. We then show that the transition to intermittency is discontinuous. Bounding the right-hand side of (\ref{secboundingF}) by (\ref{ninthbound}) with $\delta=c(r)\, y^{1/4}$, $\tau=\sqrt{y}$, and setting $K=(3T/8)^{1/4} y^{1/8}$ for a reason that will become clear shortly (see the remark below (\ref{reflecprinciple})), we obtain
\begin{eqnarray}\label{firstcontrolofF}
F(y)&\le& P\bigl[\, \|x(\cdot)\|_\infty\le 2\left(c(r)\, y^{1/4} +(3T/8)^{1/4} y^{1/8}\right)\, \bigr] \nonumber \\
&+&P\left(\omega_{x(\cdot)}(\sqrt{y})>(3T/8)^{1/4} y^{1/8}\right) .
 \end{eqnarray}
The first term on the right-hand side of (\ref{firstcontrolofF}) is bounded by the small-ball estimate \cite{Chung1948,LS2001,Fatalov2003}
$$
P(\|x(\cdot)\|_\infty\le\theta)\le \frac{4}{\pi}\, \exp\left(-\frac{\pi^2 T}{8\theta^2}\right) ,
$$
which yields
\begin{equation}\label{smallballest1}
P\bigl[\, \|x(\cdot)\|_\infty\le 2\left(c(r)\, y^{1/4} +(3T/8)^{1/4} y^{1/8}\right)\, \bigr]\le\frac{4}{\pi}
\, \exp\left(-\frac{\pi^2 T}{32\left(c(r)\, y^{1/4} +(3T/8)^{1/4} y^{1/8}\right)^2}\right) .
\end{equation}
From
$$
c(r)\, y^{1/4} +(3T/8)^{1/4} y^{1/8}\sim (3T/8)^{1/4} y^{1/8} \qquad (y\to 0),
$$
it follows that for every $\alpha>0$, there exists $y_2>0$ such that for all $y\le y_2$,
$$
\left\vert\frac{c(r)\, y^{1/4} +(3T/8)^{1/4} y^{1/8}}{(3T/8)^{1/4} y^{1/8}}-1\right\vert\le\alpha .
$$
In particular, taking $\alpha=\pi -1$ one has
$$
c(r)\, y^{1/4} +(3T/8)^{1/4} y^{1/8}\le\pi (3T/8)^{1/4} y^{1/8},
$$
and, for every $y\le y_2$, (\ref{smallballest1}) gives the estimate
\begin{equation}\label{smallballest2}
P\bigl[\, \|x(\cdot)\|_\infty\le 2\left(c(r)\, y^{1/4} +(3T/8)^{1/4} y^{1/8}\right)\, \bigr]\le\frac{4}{\pi}
\, \exp\left(-\frac{\sqrt{2T/3}}{16\, y^{1/4}}\right) .
\end{equation}
It is proved in appendix \ref{app2} that the second term on the right-hand side of (\ref{firstcontrolofF}) is bounded by the inequality
$$
P\left(\omega_{x(\cdot)}(\tau)>\theta\right)\le
\frac{8\sqrt{3}}{\sqrt{\pi}}\, \frac{T}{\theta\sqrt{\tau}}\, \exp\left(-\frac{\theta^2}{12\tau}\right),
$$
which gives
\begin{equation}\label{reflecprinciple}
P\left(\omega_{x(\cdot)}(\sqrt{y})>T^{1/4} y^{1/8}/\sqrt{2}\right) \le
\frac{16}{\sqrt{\pi}}\, \left(\frac{3}{2}\right)^{1/4}\, \frac{T^{3/4}}{y^{3/8}}\, \exp\left(-\frac{\sqrt{2T/3}}{16\, y^{1/4}}\right) .
\end{equation}
The fact that we have the same exponential in (\ref{smallballest2}) and (\ref{reflecprinciple}) is due to our choice for $K$. Combining the estimates (\ref{smallballest2}) and (\ref{reflecprinciple}) in equation (\ref{firstcontrolofF}), we obtain the desired control of $F(y)$ near $y=0$ as
\begin{equation}\label{finalcontrolofF}
F(y)\le\frac{4}{\pi}\, \left(1+6\sqrt{\pi}\, \frac{(2T/3)^{3/4}}{y^{3/8}}\right)
\, \exp\left(-\frac{\sqrt{2T/3}}{16\, y^{1/4}}\right) ,
\end{equation}
which holds for all $y\le\min(y_1,y_2)$, where $y_1$ and $y_2$ are defined below equations (\ref{constantcofr}) and (\ref{smallballest1}), respectively.

According to (\ref{finalcontrolofF}), $F(y)$ goes to zero faster than any power of $y$ as $y\to 0$, and by Stieltjes integration by parts,
\begin{equation}\label{integbyparts}
\int_0^{\mu_{\rm max}}\frac{dF(y)}{\sqrt{y}}=\left.\frac{F(y)}{\sqrt{y}}\right\vert_0^{\mu_{\rm max}}
+\int_0^{\mu_{\rm max}}\frac{F(y)}{2\, y^{3/2}}\, dy <\infty .
\end{equation}
The rapid decay of $F(y)$ near $y=0$ ensures the existence of each of the two terms on the right-hand side of the equality. Consequently, $\langle\mathcal{E}(0,g_c)\rangle <\infty$, and the transition to intermittency in the diffusion-amplification problem (\ref{modifiedwithDeq}) is discontinuous for the considered class of $S$.
%
%%%%%%%%%%
%
\section{Discussion and perspectives}\label{sec6}
In this paper, we addressed the question of diffusion effects on the transition to intermittency in stochastic amplification driven by the square of a Gaussian field. We have analysed a model in which the diffusion-amplification process takes place in a random lattice of point absorbers (equation (\ref{modifiedwithDeq})). The presence of the absorbers naturally restricts the Feynman-Kac representation to Brownian paths with range not exceeding a finite cut-off set by the largest possible spacing between neighboring absorbers (see equations (\ref{quenchedamplisol}) and (\ref{latticeaverage1})). In this case and for $S$ in the class introduced in section \ref{sec2a}, we have shown that the transition to intermittency is discontinuous, in the sense that the associated order parameter is discontinuous at the transition. In contrast, the diffusion-free case exhibits a continuous transition. This comparison indicates that diffusion-induced damping of the amplification is strong enough to make $\langle\mathcal{E}(0,g_c)\rangle<\infty$, thereby changing the nature of the transition.

Extending our results to the original problem (\ref{withDeq}) considered in \cite{Mounaix2026}, in which no absorbers are present, is a natural next step. Although this problem is closely related to the one considered above, the same line of calculations cannot be followed without encountering difficulties that we are unable to resolve at the present time. What prevents us from applying our argument is that, for any $\rho>0$, $Y[x(\cdot)]=0$ can be approached arbitrarily closely by continuous paths with $R_T[x(\cdot)]>\rho$ (see the discussion at the end of section \ref{sec1b}). If this were not the case, there would exist an $r>0$ and a constant $c>0$ such that $Y[x(\cdot)]\ge c$ for all path with $R_T[x(\cdot)]>r$. The contribution of these paths to $\langle\mathcal{E}(0,g_c)\rangle$ would be finite, and proving $\langle\mathcal{E}(0,g_c)\rangle<\infty$ would then reduce to proving the finiteness of the contribution from paths with $R_T[x(\cdot)]\le r$, which is precisely the calculation carried out in this paper starting from equation (\ref{boundingE}). But there is no such $r$, and the above reasoning does not apply. In technical terms, replacing $\psi(x,t)$ with the solution to (\ref{withDeq}) in our analysis amounts to letting $r\to\infty$ throughout the calculations starting from equation (\ref{boundingE}). In particular, taking the limit $r\to\infty$ in equation (\ref{defofM}) yields $M_\infty(\delta)=0$, and the lower bound in (\ref{boundthreeY}) reduces to the trivial, useless inequality $Y[x(\cdot)]\ge 0$. The crux of the difficulty is therefore to obtain a non-trivial lower bound for $\varphi(u)$, analogous to (\ref{pointboundofphi}) for finite $r$, that would allow us to bound $Y[x(\cdot)]$ from below by a quantity independent of $\varphi$, as on the right-hand side of (\ref{boundthreeY}). In the absence of such a bound, we are left with the estimate in (\ref{boundtwoY}), which still depends on $\varphi$. Since $\varphi$ depends on $x(\cdot)$ in an unknown, hence uncontrollable, manner, this prevents us from controlling $F(y)$, at least within the framework of the present argument. For the time being, it therefore remains an open question whether the contribution of paths with arbitrarily large range might modify the small-$y$ behavior of $F(y)$ in a way that would restore a continuous transition. Resolving this question will likely require a different approach, ideally one general enough to also handle a wider class of $S$ than in section \ref{sec2a}, such as that considered in \cite{Mounaix2026}.

When the correlation function of $S$ is specified, the problem can in principle also be addressed through numerical investigations, although this remains technically challenging. Since numerical simulations necessarily involve both finite systems and a finite number of realizations, the corresponding sample average of $\mathcal{E}(0,g)$ remains finite, and $Q(g)>0$ for all $g$. The numerical signature of a discontinuous transition to intermittency is thus expected to take the form of an abrupt drop in $Q(g)$, from $0<Q(g)\lesssim 1$ to $0<Q(g)\ll 1$, over a narrow interval of $g$. To capture such a behavior unambiguously, a very large number of realizations is required to adequately sample the rare values of $\mathcal{E}(0,g)$ in the far upper tail of its distribution, which dominate its sample average in the intermittent regime. Moreover, in the vicinity of $g=g_c$, the required sampling becomes even more demanding as the average of $\mathcal{E}(0,g_c)$ is entirely controlled by the slowly varying (subdominant) amplitude of the far upper tail, which is particularly difficult to resolve numerically. In concrete terms, equation (\ref{withDeq}) must be solved numerically for each realization of $S$ and for a large system size $L$ (much larger than the correlation length of $S$). This computation must then be repeated for a very large number of realizations of $S$ in order to obtain reliable statistics of $\mathcal{E}(0,g)$. Finally, the entire procedure must be repeated for several closely spaced values of $g$ near $g_c$, which itself must be determined numerically. A brute-force implementation of this program would require a prohibitively large number of numerical solutions of equations (\ref{withDeq}). A possible way to circumvent this difficulty would be to bias the distribution of $S$ towards the realizations that dominate the sample average of $\mathcal{E}(0,g_c)$, for example by developing an importance sampling algorithm specifically tailored to this problem. Even if such an approach proved feasible -- which is far from obvious a priori -- it would still require massive numerical simulations. Undertaking such a numerical study would be of considerable interest, but it lies beyond the scope of the present paper.

In the longer run, there are various directions along which investigations could be pushed further. Of particular interest is the replacement of $\partial^2_{x^2}$ with $D\, \partial^2_{x^2}$ in equation (\ref{withDeq}), where $D\ne 0$ is a complex diffusivity with ${\rm Re}(D)\ge 0$. This generalization encompasses the physically important diffractive version of (\ref{withDeq}), when ${\rm Re}(D)=0$. We refer the interested reader to \cite{Mounaix2025,MCL2006,Mounaix2023,Mounaix2024} for insight into the highly non-trivial mathematical challenges arising in this broader setting.

Another intriguing line of research is to understand the connection between the results obtained here and in \cite{Mounaix2026} for real $D>0$ and those of the diffusion-free case $D=0$. More precisely, let $L>0$ denote the size of the system, with periodic boundary conditions at $x=\pm L/2$. Our results, which pertain to the limit $L\to\infty$ at fixed $D>0$, differ from those obtained for $D=0$ in the same limit (see section \ref{sec0}). This indicates that the two limits $L\to\infty$ and $D\to 0$ do not commute, which raises the question of how precisely the crossover between `$L\to\infty$ then $D\to 0$' and `$D\to 0$ then $L\to\infty$' occurs. Answering this question would elucidate the shift, as $D$ goes to zero, from a discontinuous transition at a critical coupling to a continuous transition at a possibly larger critical coupling (when $\sup_{x(\cdot)\in B(0,T)}\mu_1[x(\cdot)]>\mu_1[x(\cdot)\equiv 0]$).

Before closing this discussion, we briefly note that the more familiar picture in terms of ordered vs. disordered phases can be recovered by viewing the transition through the normalized field
$$
\tilde{\mathcal{E}}(x,g)=\mathcal{E}(x,g)/\langle\mathcal{E}(0,g)\rangle ,
$$
as suggested by the definition of $Q(g)$. From this perspective, the intermittency of $\mathcal{E}(x,g)$ manifests itself in the vanishing of $\tilde{\mathcal{E}}(x,g)$ almost everywhere, with normalized mass $\tilde{m}(g)=~0$, reflecting the supercritical loss of ergodicity (see proposition 3 in \cite{Mounaix2026}). As $g$ decreases from the supercritical regime, the transition at $g=g_c$ is marked by the emergence of a strictly positive field $\tilde{\mathcal{E}}(x,g)$, whose mass satisfies $\tilde{m}(g)=1$ by ergodicity. Defining the degree of order in terms of how evenly mass is apportioned among points in space, the subcritical regime can thus be viewed as an ordered phase, with $\tilde{\mathcal{E}}(x,g)$ playing the role of a local order-parameter field: vanishing almost everywhere in the supercritical regime, where almost no point contributes to the mass, and strictly positive in the subcritical regime, where mass is distributed relatively evenly across space, albeit with local inhomogeneities (and with strict uniformity at $g=0$).

In conclusion, this work represents a first step toward a comprehensive study of the effects of propagation on the transition to intermittency in stochastic amplifier models. While our analysis provides a complete understanding of the transition for problem (\ref{modifiedwithDeq}), much remains to be understood, with several non-trivial open questions providing natural avenues for future research. We hope that the results presented here will stimulate further work on these topics in both statistical and mathematical physics.
%
%%%%%%%%%%%%%%%%%%%%
%
\acknowledgments{The author warmly thanks St\'ephane Munier and Gr\'egory Schehr for their interest and valuable advice about the manuscript.}
%
%%%%%%%%%%%%%%%%%%%%
%
\appendix
\section{Bounding $\bm{\langle\exp(-\int_0^T V(x(\tau))\, d\tau)\rangle_{a.p.}}$ by $\bm{1_{R_T[x(\cdot)]\le r}}$}\label{app1}
In this appendix, we prove the inequality (\ref{boundingabsav}). For each path $x(\cdot)$ in $B(x,T)$, we define
$$
\begin{array}{l}
m=\inf_{0\le t\le T}x(t), \\
M=\sup_{0\le t\le T}x(t),
\end{array}
$$
and we write $R_T\equiv M-m$ as shorthand for $R_T[x(\cdot)]$. If $R_T\ge r$, there is at least one absorber in $[m,M]$ and since $x(\cdot)$ is continuous, it follows that
\begin{equation}\label{latticeaverage1}
\left\langle {\rm e}^{ -\int_0^T V(x(\tau))\, d\tau}\right\rangle_{a.p.} =0 \qquad (R_T\ge r).
\end{equation}
On the other hand, if $R_T<r$, then
\begin{eqnarray}\label{latticeaverage2}
\left\langle {\rm e}^{ -\int_0^T V(x(\tau))\, d\tau}\right\rangle_{a.p.} &=&
{\rm Prob.}({\rm no\ absorber\ in}\ [m,M]) \nonumber \\
&=&{\rm Prob.}({\rm no\ absorber\ in}\ [0,R_T]) \nonumber \\
&=&\int_{R_T}^r\left(1-\frac{R_T}{d_0}\right)\, d\pi(d_0) \qquad (R_T<r),
\end{eqnarray}
where we have used translational invariance of the absorber position statistics (second equality), and the uniform distribution of $x_0$, given $d_0$ (third equality). Equations (\ref{latticeaverage1}) and (\ref{latticeaverage2}) show that only paths satisfying $R_T[x(\cdot)]<r$ contribute to $\mathcal{E}(x,g)$. Embedding the diffusion-amplification process in the random lattice of point absorbers defined in section \ref{sec1a} therefore introduces a natural cut-off on $R_T[x(\cdot)]$. Since
$$
\int_{R_T}^r\left(1-\frac{R_T}{d_0}\right)\, d\pi(d_0)\le\pi(R_T<d_0\le r)\le 1,
$$
one has
$$
\left\langle {\rm e}^{ -\int_0^T V(x(\tau))\, d\tau}\right\rangle_{a.p.}\le
\bm{1}_{R_T[x(\cdot)]\le r},
$$
which completes the proof of equation (\ref{boundingabsav}).
%
%%%%%%%%%%%%%%%%%%%%
%
\section{Proof of $\bm{\mu_2[x(\cdot)]/\mu_1[x(\cdot)]\le 1-C(r,T)}$}\label{appSG}
In this appendix, we prove the inequality (\ref{spectralgap}) by using the Doob $h$-transform \cite{Doob1957} and the associated Doeblin minorization condition \cite{MT2009}.

For each $x(\cdot)$ in $B_r(0,T)$, let $\varphi$ denote the (strictly) positive eigenfunction of the kernel $C(x(s)-x(u),s-u)$ associated with the eigenvalue $\mu_1[x(\cdot)]$, normalized to $\|\varphi\|_2=1$. Let $P_{x(\cdot)}$ be the Doob $h$-transform of $C(x(s)-x(u),s-u)$ for $h=\varphi$, defined by
\begin{equation}\label{eqSG1}
\left(P_{x(\cdot)}f\right)(s)
=\int_0^T f(u)P_{x(\cdot)}(s,du),
\end{equation}
where
\begin{equation}\label{eqSG2}
P_{x(\cdot)}(s,du)=\frac{C(x(s)-x(u),s-u)\, \varphi(u)}{\mu_1[x(\cdot)]\, \varphi(s)}\, du,
\end{equation}
is the corresponding Markov transition kernel \cite{Doob1957}. From the eigenvalue equation and the upper bound $C(x,t)\le 1$, we have
\begin{eqnarray*}
\mu_1[x(\cdot)]\, \varphi(s) &=&\int_0^T C(x(s)-x(u),s-u)\, \varphi(u)\, du \\
&\le&\int_0^T \varphi(u)\, du.
\end{eqnarray*}
Therefore,
$$
\frac{1}{\mu_1[x(\cdot)]\, \varphi(s)}\ge\frac{1}{\int_0^T \varphi(u)\, du}.
$$
Substituting this estimate into (\ref{eqSG2}) and using $C(x(s)-x(u),s-u)\ge C(r,T)$ (which follows from condition \ref{condition3} in section \ref{sec2a}), we obtain the Doeblin minorization condition \cite{MT2009},
\begin{equation}\label{eqSG3}
P_{x(\cdot)}(s,du)\ge C(r,T)\, \nu_{x(\cdot)}(du),
\end{equation}
with probability measure
$$
\nu_{x(\cdot)}(du)=\frac{\varphi(u)}{\int_0^T \varphi(u)\, du}\, du,
$$
and minorization constant $C(r,T)\le 1$.

We denote by $\operatorname{osc}(f)=\sup_{0\le s\le T}f(s)-\inf_{0\le s\le T}f(s)$ the oscillation of $f$. From (\ref{eqSG3}), it follows that there exists a Markov transition kernel $Q_{x(\cdot)}(s,du)$ such that
\begin{equation}\label{eqSG4}
P_{x(\cdot)}(s,du)= C(r,T)\, \nu_{x(\cdot)}(du)+\left[ 1-C(r,T)\right]\, Q_{x(\cdot)}(s,du).
\end{equation}
(See, e.g., the splitting construction in \cite{MT2009}.) The first term on the right-hand side of (\ref{eqSG4}) is independent of $s$, hence
\begin{equation}\label{eqSG5}
\operatorname{osc}\left( P_{x(\cdot)}f\right) = \left[ 1-C(r,T)\right]\, \operatorname{osc}\left( Q_{x(\cdot)}f\right) .
\end{equation}
Since $Q_{x(\cdot)}(s,du)$ is a Markov kernel, $(Q_{x(\cdot)}f)(s)$ is an average of values of $f$. Therefore,
$$
\inf_{0\le t\le T}f(t)\le\left(Q_{x(\cdot)}f\right)(s)\le\sup_{0\le t\le T}f(t),
$$
and equation (\ref{eqSG5}) yields
\begin{equation}\label{eqSG6}
\operatorname{osc}\left( P_{x(\cdot)}f\right) \le \left[ 1-C(r,T)\right]\, \operatorname{osc}(f) .
\end{equation}
Let $\phi$ be an eigenfunction of the kernel $C(x(s)-x(u),s-u)$ associated with the eigenvalue $\mu_2[x(\cdot)]$, and define $f_2=\phi/\varphi$. Equations (\ref{eqSG1}) and (\ref{eqSG2}) yield
$$
\left( P_{x(\cdot)}f_2\right)(s)=\frac{\mu_2[x(\cdot)]}{\mu_1[x(\cdot)]}\, f_2(s).
$$
Hence,
\begin{equation}\label{eqSG7}
\operatorname{osc}\left( P_{x(\cdot)}f_2\right)=\frac{\mu_2[x(\cdot)]}{\mu_1[x(\cdot)]}\, \operatorname{osc}(f_2).
\end{equation}
Substituting (\ref{eqSG7}) into the oscillation contraction (\ref{eqSG6}) with $f=f_2$, we obtain,
\begin{equation}\label{eqSG8}
\frac{\mu_2[x(\cdot)]}{\mu_1[x(\cdot)]}\, \operatorname{osc}(f_2)
 \le \left[ 1-C(r,T)\right]\, \operatorname{osc}(f_2).
 \end{equation}
 Since $f_2$ is not constant, its oscillation is strictly positive and (\ref{eqSG8}) implies (\ref{spectralgap}):
$$
\frac{\mu_2[x(\cdot)]}{\mu_1[x(\cdot)]}\le 1-C(r,T).
$$
%
%%%%%%%%%%%%%%%%%%%%
%
\section{Bounding $\bm{P\left(\omega_{x(\cdot)}(\tau)>\theta\right)}$ via the reflection principle}\label{app2}
Since $x(t)$ and $x(T-t)$ have the same modulus of continuity on $[0,T]$ (equation (\ref{defofomega})), we take $x(\cdot)$ to be a standard Brownian motion on $[0,\infty)$ starting at $x(0)=0$ throughout this appendix. Consider the $N+1$ intervals
$$
I_j=[j\tau/2, j\tau/2+3\tau/2],\qquad j=0,\, 1,\, 2,\, \ldots ,\, N
$$
where $\tau<T$ and $N=\lceil 2T/\tau -3\rceil\ge 0$, whose union covers the interval $[0,T]$. By construction, every pair $s,u\in [0,T]$ satisfying $\vert s-u\vert\le\tau$ is contained in at least one of the intervals $I_j$. Therefore, if $\omega_{x(\cdot)}(\tau)>\theta$ for some $\theta>0$, then
$$
R_j=\sup_{s,u\in I_j}\vert x(s)-x(u)\vert
$$
exceeds $\theta$ for at least one $j$. In other words,
$$
\left\lbrace\omega_{x(\cdot)}(\tau)>\theta\right\rbrace\subseteq
\bigcup_{j=0}^N \left\lbrace R_j >\theta\right\rbrace ,
$$
and hence, by the union bound,
$$
P\left(\omega_{x(\cdot)}(\tau)>\theta\right)\le\sum_{j=0}^N P\left(R_j >\theta\right).
$$
By the stationarity of Brownian increments, $R_j$ has the same distribution as
$$
R_{3\tau/2}=\sup_{0\le s,u\le 3\tau/2}\vert x(s)-x(u)\vert.
$$
Thus,
\begin{equation}\label{eq1app2}
P\left(\omega_{x(\cdot)}(\tau)>\theta\right)\le\frac{2T}{\tau}\, P\left(R_{3\tau/2} >\theta\right),
\end{equation}
where we have used $N+1<2T/\tau -1<2T/\tau$. Since
$$
R_{3\tau/2}\le 2\sup_{0\le t\le 3\tau/2}\vert x(t)\vert,
$$
we have
$$
P\left(R_{3\tau/2} >\theta\right)\le P\left(\sup_{0\le t\le 3\tau/2}\vert x(t)\vert >\frac{\theta}{2}\right).
$$
By symmetry,
$$
P\left(\sup_{0\le t\le 3\tau/2}\vert x(t)\vert >\frac{\theta}{2}\right)
\le 2\, P\left(\sup_{0\le t\le 3\tau/2}x(t) >\frac{\theta}{2}\right) ,
$$
and hence,
\begin{equation}\label{intermediateineq}
P\left(R_{3\tau/2} >\theta\right)\le 2\, P\left(\sup_{0\le t\le 3\tau/2}x(t) >\frac{\theta}{2}\right).
\end{equation}
Combining (\ref{intermediateineq}) with (\ref{eq1app2}) gives
\begin{equation}\label{eq2app2}
P\left(\omega_{x(\cdot)}(\tau)>\theta\right)\le\frac{4T}{\tau}\, P\left(\sup_{0\le t\le 3\tau/2}x(t) >\frac{\theta}{2}\right).
\end{equation}
Finally, the probability on the right-hand side of (\ref{eq2app2}) can be bounded using the reflection principle \cite{Varadhan2007,MP2010}:
$$
P\left(\sup_{0\le t\le 3\tau/2}x(t)>\frac{\theta}{2}\right)\le
\frac{2\sqrt{3\tau}}{\theta\sqrt{\pi}}\, \exp\left(-\frac{\theta^2}{12\tau}\right) ,
$$
which yields the desired estimate
\begin{equation}\label{eq3app2}
P\left(\omega_{x(\cdot)}(\tau)>\theta\right)\le
\frac{8\sqrt{3}}{\sqrt{\pi}}\, \frac{T}{\theta\sqrt{\tau}}\, \exp\left(-\frac{\theta^2}{12\tau}\right).
\end{equation}
The bound in (\ref{eq3app2}) is not optimal, but it is sufficient for our purposes.
%
%%%%%%%%%%%%%%%%%%%%
%

%
%%%%%%%%%%%%%%%%%%%%
%
\end{document}